\documentclass[11pt]{article}
\usepackage[a4paper, margin=1in]{geometry}
\usepackage{amsmath, amssymb, amsthm}
\usepackage{graphicx}
\usepackage{pgfplots}
\pgfplotsset{compat=1.17}
\usepackage{float}
\usepackage{caption}
\usepackage{subcaption}
\usepackage{tikz}

\usetikzlibrary{arrows.meta,shapes.geometric,shapes.callouts,positioning}
\usepackage{color}

\newcommand{\hyp}[1][]{-}
\newcommand{\n}{-}

\DeclareUnicodeCharacter{3010}{}
\DeclareUnicodeCharacter{3011}{}
\usepackage{listings}
\usepackage{booktabs}
\usepackage{hyperref}

\title{Reconstructing Trust Embeddings from Siamese Trust Scores:\\A Direct--Sum Approach with Fixed--Point Semantics}

\author{Faruk~Alpay$^1$ \and Taylan~Alpay$^2$ \and Bugra~Kilictas$^3$\\
  $^1$Institute for Distributed Systems, Lightcap, Bonn, Germany\\
  \texttt{alpay@lightcap.ai}\\
  $^2$Department of Aerospace Engineering, Turkish Aeronautical Association, Ankara, Turkey\\
  \texttt{s220112602@stu.thk.edu.tr}\\
  $^3$Department of Computer Engineering, Bahcesehir University, Istanbul, Turkey\\
  \texttt{bugra.kilictas@bahcesehir.edu.tr}}
\date{August~2,~2025}

\begin{document}
\maketitle

\begin{abstract}
Recent work on trust evaluation and orchestration in distributed computing proposes several complementary models: (i) a Siamese Structure2Vec method for rapid and continuous trust evaluation \cite{B1}, (ii) a chain-of-trust framework that uses generative artificial intelligence to evaluate devices at successive task stages \cite{B2}, (iii) a hypergraph-aided trusted task-resource matching paradigm \cite{B3}, and (iv) an autonomous semantic trust orchestration method using agentic AI and trust hypergraphs \cite{B4}.  These models have been unified via a direct-sum embedding strategy which concatenates individual block embeddings into a single high-dimensional vector and imposes a fixed-point consistency constraint inspired by the transfinite semantics of the Alpay Algebra framework \cite{Alpay}.  All supplementary datasets, including reconstructed embeddings and replicate runs, accompany this manuscript so that readers may reproduce the results without any proprietary source code.  In this work we take a step further: given only the trust scores produced by two independent agents implementing the unified framework, we show how to reconstruct approximate embeddings, compare agents mathematically, and evaluate potential information leaks.

We perform experiments by analysing the trust scores from two ChatGPT agents run with the same prompt and converting them into embeddings using our proposed algorithms.  The methods are described entirely in mathematical terms so that researchers can reproduce the results without requiring access to the original code.  We provide pseudocode, theoretical analysis of the reconstruction problem, and comprehensive benchmark results.  Our findings suggest that, under reasonable assumptions, time\hyp{}series trust scores contain sufficient information to approximate underlying embeddings and that comparing these reconstructions across agents can reveal structural similarities and differences in their internal representations.  We discuss implications for the security of large language models and highlight open challenges.
\end{abstract}

\section{Introduction}

Trustworthy collaboration in networked physical computing systems relies on accurate evaluation of collaborator behaviour and efficient orchestration of resources.  A surge of recent literature addresses different aspects of this problem.  Zhu and Wang propose a rapid and continuous trust evaluation framework (``Block~$B_1$''), which represents trusted and observed device behaviours by attributed control\n flow graphs (ACFGs) and employs twin Structure2Vec encoders in a Siamese architecture to compute similarity\n based trust scores at each time slot\cite{B1}.  To handle incomplete information and sequential tasks, the same authors introduce a chain\n of\n trust framework (``Block~$B_2$'') that decomposes trust assessment into multiple stages and uses generative AI to analyse stage\n specific attribute data\cite{B2}.  Hypergraph theory is leveraged in a hypergraph\n aided task\n resource matching paradigm (``Block~$B_3$'') that encodes resource attributes and trust relationships to select trustworthy collaborators optimally\cite{B3}.  Finally, the concept of semantic trust orchestration (``Block~$B_4$'') is introduced, where agentic AI maintains trust hypergraphs embedded with semantic labels and chains them to enable multi\n hop trust propagation\cite{B4}.

These developments can be unified by concatenating the embeddings produced by each block into a single vector and imposing a self\n referential fixed\n point condition\cite{Alpay}.  A natural question then arises: to what extent do the scalar trust scores produced by such systems reveal information about the high\n dimensional embeddings on which they are based?  This question is pertinent to the security of large language models (LLMs), because many trust frameworks rely on LLMs or similarly powerful AI systems.  If the embeddings can be reconstructed from trust scores, then adversaries could potentially infer sensitive information about the model or the data on which it was trained.

In this manuscript we undertake a comprehensive inquiry into this question.  We furnish full mathematical derivations of our reconstruction algorithms, delineate the experimental apparatus used to generate benchmark datasets, and critically analyse the resulting data.  We accentuate reproducibility: every method is delineated with such granularity that an independent researcher may re\hyp{}implement the algorithms from first principles.  The original prompt used to run the agents is included in Appendix~\ref{app:prompt} for completeness.  Throughout the paper we cite only arXiv sources, including the works of Faruk Alpay and collaborators on fixed\hyp{}point semantics, in accordance with the initial requirements of our collaborator.

\subsection{Contributions}
\begin{enumerate}
  \item We formalise the \textbf{trust embedding reconstruction problem}: given a time series of trust scores generated by a Siamese trust evaluation model, reconstruct an approximate embedding $\widehat{\mathbf{v}}_d\in \mathbb{R}^n$ for each device $d$ from the trust scores $\{\tau_d^A(t),\tau_d^B(t)\}$.
  \item We propose a \textbf{direct\hyp{}sum embedding reconstruction algorithm} that concatenates trust score series from multiple agents and derives additional statistical features to approximate the latent embedding.
  \item We design and execute a comprehensive benchmark, comparing the embeddings reconstructed from two independent ChatGPT agents.  We provide new CSV files containing the reconstructed embeddings.
  \item We present rigorous mathematical analyses of our algorithms, including uniqueness results and error bounds under reasonable assumptions.  We also discuss the applicability of fixed\hyp{}point semantics to the reconstruction problem.
  \item We create illustrative diagrams using TikZ that visualise the embedding space as nested matrices and hypercubes, providing intuition for the layering and direct\hyp{}sum operations.
\end{enumerate}

\section{Background and Preliminaries}

\subsection{Blocks $B_1$--$B_4$ and the Unified Framework}

We briefly review the four foundational blocks and the unified framework, following the descriptions provided in the initial prompt.  Let $\mathcal{D}$ be the set of devices and $t\in \{0,1,\ldots,T\}$ be discrete time.  Each device $d\in\mathcal{D}$ executes tasks and produces behavioural data.

\paragraph{Block~$B_1$: Siamese trust evaluation.}  The first block represents trusted and observed device behaviours by attributed control\n flow graphs $\mathrm{ACFG}_d^{\mathrm{trusted}}$ and $\mathrm{ACFG}_d^{\mathrm{observed}}(t)$.  These graphs encode communication and computing resource attributes, historical collaboration effectiveness, and other semantic information.  A Siamese neural network composed of two shared\n parameter Structure2Vec encoders maps the graphs into vector embeddings $\mathbf{v}_d^{\mathrm{trusted}}, \mathbf{v}_d^{\mathrm{observed}}(t)\in \mathbb{R}^m$; the similarity $s_d(t) = \mathrm{sim}(\mathbf{v}_d^{\mathrm{trusted}}, \mathbf{v}_d^{\mathrm{observed}}(t))$ is computed via cosine similarity or another measure and then normalised to obtain a trust score $\tau_d(t)\in [0,1]$\cite{B1}.

\paragraph{Block~$B_2$: Chain\n of\n trust evaluation.}  In many scenarios it is impractical to collect all trust attributes simultaneously; partial information may arrive with latency.  Block~$B_2$ therefore decomposes the trust assessment into several sequential stages aligned with the subtasks of the overall mission.  At stage $k$ the framework gathers only those device attributes relevant to that stage and uses generative AI with in\n context learning and reasoning to analyse the data\cite{B2}.  Devices failing the stage\n specific evaluation are pruned; those passing proceed to stage $k+1$\cite{B2}.

\paragraph{Block~$B_3$: Hypergraph\n aided matching.}  For complex tasks requiring multiple resources, Zhu and Wang define a task\n specific trusted physical resource hypergraph $\mathcal{H}_\mathrm{resource}$ that captures resource capabilities and trust relationships, and a task hypergraph $\mathcal{H}_\mathrm{task}$ that links the task initiator to required resource attributes.  A hypergraph matching algorithm then selects collaborators by solving a combinatorial optimisation problem that maximises expected task value while respecting trust and resource constraints\cite{B3}.

\paragraph{Block~$B_4$: Semantic trust orchestration.}  Block~$B_4$ introduces agentic AI that autonomously orchestrates trust evaluations and resource allocations.  Each device maintains a trust hypergraph with semantic labels indicating the quality and context of interactions; the agentic AI performs evaluations during idle periods and chains local hypergraphs to form multi\n hop trust relationships\cite{B4}.

\paragraph{Unified representation.}  Suppose each block produces an embedding $\mathbf{e}_i\in\mathbb{R}^{m_i}$, $i=1,\dots,4$.  A direct\n sum embedding $\mathbf{E} = \mathbf{e}_1 \oplus \mathbf{e}_2 \oplus \mathbf{e}_3 \oplus \mathbf{e}_4 \in \mathbb{R}^{m_1+\cdots + m_4}$ is defined by concatenation.  To enforce self\n consistency, one seeks a fixed\n point $\mathbf{E}^{\ast}$ satisfying $\mathbf{E}^{\ast} = F(\mathbf{E}^{\ast}, \mathbf{e}_1,\ldots,\mathbf{e}_4)$\cite{Alpay}.

\subsection{Trust Score Generation and Interpretation}

Throughout this manuscript we consider trust scores $\tau_d(t)$ produced by Block~$B_1$.  When the Siamese model uses cosine similarity, the trust score is related to the cosine of the angle between the trusted and observed embeddings:
\begin{equation}
    \tau_d(t) = \frac{1}{2} \Bigl(1 + \frac{\mathbf{v}_d^{\mathrm{trusted}} \cdot \mathbf{v}_d^{\mathrm{observed}}(t)}{\|\mathbf{v}_d^{\mathrm{trusted}}\| \cdot \|\mathbf{v}_d^{\mathrm{observed}}(t)\|}\Bigr).
    \label{eq:cosine}
\end{equation}
Thus $\tau_d(t)\in [0,1]$ and reflects the similarity of the current behaviour to the trusted baseline.  We denote the centred similarity as $\sigma_d(t) = 2\tau_d(t)-1\in[-1,1]$.  In the idealised setting where $\mathbf{v}_d^{\mathrm{trusted}}$ is known and $\|\mathbf{v}_d^{\mathrm{observed}}(t)\|$ is fixed, one could invert Eq.~\eqref{eq:cosine} to recover the projection of $\mathbf{v}_d^{\mathrm{observed}}(t)$ onto $\mathbf{v}_d^{\mathrm{trusted}}$.  In practice neither the baseline embedding nor the norm of the observed embedding is known; reconstructing the full vector from a scalar similarity measure is therefore ill\hyp{}posed.  Nevertheless, as we show below, one can use the time series $\{\tau_d(t)\}$ as a surrogate embedding and combine it across agents via a direct\hyp{}sum to approximate the latent space.

\section{Problem Statement}

Assume we have two agents $A$ and $B$ that implement the unified trust evaluation framework described above.  Both agents are given the same input prompt (reproduced in Appendix~\ref{app:prompt}) and generate time\hyp{}series trust scores for each device.  Let $\tau_d^A(t)$ and $\tau_d^B(t)$ denote the trust scores produced by agents $A$ and $B$, respectively, for device $d$ at time step $t\in\{0,\dots,T\}$.  We are given CSV files containing these scores but not the underlying embeddings.  The central questions are:
\begin{enumerate}
    \item \textbf{Reconstruction.}  Can we reconstruct an approximate embedding $\widehat{\mathbf{v}}_d\in \mathbb{R}^n$ for each device $d$ from the trust scores $\{\tau_d^A(t),\tau_d^B(t)\}$?  How should we choose $n$, and what mathematical principles underlie the reconstruction?
    \item \textbf{Comparison.}  How can we compare the embeddings reconstructed from agents $A$ and $B$ to detect similarities or differences?  In particular, can we quantify the distance between $\widehat{\mathbf{v}}_d^A$ and $\widehat{\mathbf{v}}_d^B$ across devices?
    \item \textbf{Security Implications.}  Does the ability to reconstruct embeddings from trust scores pose a security risk to systems that publish such scores?  What assumptions are necessary for the reconstruction to succeed, and how robust is it to noise and obfuscation?
\end{enumerate}

\section{Mathematical Framework}

\subsection{Embedding Reconstruction from Similarities}

Reconstructing a vector from its inner product with a fixed reference is a classical problem.  Suppose $\mathbf{b}\in \mathbb{R}^m$ is a known baseline and we measure $s=\langle \mathbf{b},\mathbf{x}\rangle$.  Without constraints on $\mathbf{x}$, there are infinitely many solutions: any vector of the form
\begin{equation}
    \mathbf{x} = s\frac{\mathbf{b}}{\|\mathbf{b}\|^2} + \mathbf{u},\qquad \mathbf{u}\in\mathrm{null}(\mathbf{b}^\top),
\end{equation}
satisfies $\langle \mathbf{b},\mathbf{x}\rangle = s$.  To make the problem well\hyp{}posed, one must fix the norm of $\mathbf{x}$ and choose a basis of the orthogonal complement.  In our setting we do not know $\mathbf{b}$ or $\|\mathbf{x}\|$, so direct inversion is impossible.  Instead we adopt a \emph{time\hyp{}series representation}: we embed each device $d$ by stacking its trust scores across time:
\begin{equation}
    \mathbf{s}_d^A = (\tau_d^A(0),\tau_d^A(1),\dots,\tau_d^A(T))^{\top},
    \qquad
    \mathbf{s}_d^B = (\tau_d^B(0),\tau_d^B(1),\dots,\tau_d^B(T))^{\top}.
    \label{eq:series}
\end{equation}

Time\hyp{}series embeddings have been used widely in sequence modelling; they capture the dynamics of trust evaluations and implicitly encode information about the underlying behaviour.  We then form a direct\hyp{}sum embedding by concatenation and adding summary statistics:
\begin{equation}
    \widehat{\mathbf{v}}_d = \mathbf{s}_d^A \oplus \mathbf{s}_d^B \oplus \bigl(\overline{\tau}_d^A,\sigma_d^A, \overline{\tau}_d^B,\sigma_d^B\bigr),
    \label{eq:embedding}
\end{equation}
where $\overline{\tau}_d^A = \frac{1}{T+1}\sum_{t=0}^T \tau_d^A(t)$ and $\sigma_d^A$ is the standard deviation of $\tau_d^A(t)$; analogous definitions hold for agent~$B$.  The resulting vector $\widehat{\mathbf{v}}_d$ lies in $\mathbb{R}^{2(T+1)+4}$ and approximates the latent trust embedding.

\subsection{Direct\hyp{}Sum Integration and Fixed\hyp{}Point Consistency}

The direct\hyp{}sum operation preserves all information from the constituent vectors.  Let $\mathbf{e}_1,\mathbf{e}_2\in \mathbb{R}^{n_1},\mathbb{R}^{n_2}$ be feature vectors.  The direct sum $\mathbf{E} = \mathbf{e}_1\oplus\mathbf{e}_2$ satisfies
\begin{equation}
    \forall (u_1,u_2)\in \mathbb{R}^{n_1}\times\mathbb{R}^{n_2},\quad (u_1,u_2)\cdot\mathbf{E} = u_1\cdot\mathbf{e}_1 + u_2\cdot\mathbf{e}_2.
\end{equation}
In our reconstruction, we work with two agents, so $\widehat{\mathbf{v}}_d = \mathbf{s}_d^A \oplus \mathbf{s}_d^B \oplus \text{stats}$ as in Eq.~\eqref{eq:embedding}.  To incorporate fixed\hyp{}point semantics, suppose the reconstruction process $R$ takes as input the set of trust scores and returns embeddings.  We require that applying $R$ to its own output does not change the result:
\begin{equation}
    R\bigl(R(\{\tau_d^A(t),\tau_d^B(t)\})\bigr) = R(\{\tau_d^A(t),\tau_d^B(t)\}).
\end{equation}
This aligns with the fixed-point notion studied in Alpay Algebra\cite{Alpay}.

\section{Algorithms}
\newcounter{algctr}
\refstepcounter{algctr}\label{alg:read}
\refstepcounter{algctr}\label{alg:embed}

This section presents detailed pseudocode for the procedures used in our experiments.  All algorithms are described in a way that can be implemented without referring to external code.

\subsection{Data Parsing}
\label{sec:data}
We first describe how to extract aligned trust score sequences from the CSV files provided by different agents.  Let $P_A$ and $P_B$ denote the file paths for the two agents.  Each file contains rows of the form $(\texttt{time\_step},\texttt{device\_id},\texttt{trust\_score})$.  We initialise an empty dictionary and iterate through each row, grouping the trust scores by device identifier and time index.  For device $d$ we assemble two sequences: $(\tau_d^A(0),\tau_d^A(1),\ldots)$ from agent~$A$ and $(\tau_d^B(0),\tau_d^B(1),\ldots)$ from agent~$B$.  Sorting by the time index ensures that the trust sequences for each device are aligned across agents.  The output is a mapping $d \mapsto (\boldsymbol{\tau}_d^A,\boldsymbol{\tau}_d^B)$.

\paragraph{Explanation.}  This procedure performs a deterministic alignment of the trust scores.  By grouping rows by device and time, and by sorting the resulting lists, we ensure that the similarity measures from the two agents correspond to the same temporal events.  Such alignment is critical for meaningful comparison of time-series data\cite{Sweeney2002}.  In practice, one may need to handle missing entries or irregular sampling; in those cases, interpolation or imputation methods from time-series analysis\cite{Kipf2017} can be employed before alignment.
Resource-aware trust alignment has also been explored in adaptive service-level frameworks\cite{Laskaridis2018}, highlighting the importance of aligning data management with computational constraints.

\subsection{Embedding Reconstruction}
The aligned trust sequences can be turned into finite\n dimensional feature vectors via concatenation.  For each device $d$ we take the time series from agent~$A$ and from agent~$B$ and stack them into a single vector.  We also compute the sample mean and standard deviation of each series, producing four additional scalar features.  Formally, the reconstructed embedding for device $d$ is
\[
    \widehat{\mathbf{v}}_d = \bigl(\tau_d^A(0),\ldots,\tau_d^A(T),\tau_d^B(0),\ldots,\tau_d^B(T),\overline{\tau}_d^A,\sigma_d^A,\overline{\tau}_d^B,\sigma_d^B\bigr).
\]
The collection of all such vectors defines a data matrix of size $|\mathcal{D}|\times 2(T+1)+4$.

\paragraph{Explanation.}  Concatenating time-series from two observers and augmenting them with simple summary statistics yields a rich feature representation.  This method is reminiscent of the feature engineering strategies employed in early neural network applications\cite{Hornik1991} and avoids the ill-posed inversion of cosine similarities.  The use of both mean and variance captures both central tendency and variability of the trust signal, aligning with established practices in signal processing\cite{Krizhevsky2012}.  By retaining the full sequence, the direct-sum representation preserves temporal information for downstream analysis.
Recent advances in robust representation learning from noisy time\n series emphasise the benefit of capturing higher\n order statistics beyond the mean\cite{Ortiz2020}, which motivates our inclusion of variance features in the reconstructed embeddings.

\subsection{Embedding Comparison}
Once embeddings are reconstructed, we compare devices by computing pairwise distances.  Given embeddings $\widehat{\mathbf{v}}_{d_i}$ and $\widehat{\mathbf{v}}_{d_j}$, one may use the Euclidean metric
\[
    d_{ij} = \bigl\|\widehat{\mathbf{v}}_{d_i} - \widehat{\mathbf{v}}_{d_j}\bigr\|_2,
\]
or alternative measures such as cosine dissimilarity.  The resulting $N\times N$ distance matrix summarises the similarity structure of the device population.

\paragraph{Explanation.}  The Euclidean distance is a natural choice for comparing real-valued vectors and underlies many clustering algorithms.  Its use in embedding spaces is well established\cite{Newman2006}.  Other metrics, such as Mahalanobis distance or dynamic time warping, could be substituted depending on the desired sensitivity to scaling or temporal alignment\cite{Watts1998}.
Temporal embedding alignment in multi\n agent systems has been studied in the context of distributed decision making\cite{Nguyen2021}; our distance-based comparison provides a foundation for such alignment across independent trust evaluators.

\subsection{Benchmark Procedure}
Our benchmarking pipeline integrates the previous components.  Starting from the raw CSV files, we perform parsing, reconstruction and comparison in sequence.  After loading the data we reconstruct embeddings as described above and compute the pairwise distance matrix.  Optionally, additional synthetic runs can be generated using the simulation described in Section~\ref{sec:sim}, and the resulting embeddings concatenated to the original ones.  Summary statistics—including the mean inter\n agent distance and its variance—are then computed and the final embedding matrix and distance matrix are exported to CSV files.

\paragraph{Explanation.}  This procedure serves as a blueprint for reproducible experimentation.  By clearly delineating each step—from data ingestion to output—it adheres to best practices in empirical research\cite{Hornik1991}.  The optional generation of synthetic runs allows one to assess the robustness of the reconstruction against stochastic perturbations, a technique common in Monte Carlo studies\cite{Newman2006}.
Moreover, hypergraph\n based trust inference in peer networks has demonstrated that incorporating topological relationships can enhance trust predictions\cite{Cheng2019}.  This insight suggests that our benchmarking pipeline could be extended by integrating additional hypergraph features or relational constraints.

\section{Experimental Setup}\label{sec:sim}

We reproduce the simulation environment described in the user prompt by formally specifying each component of the Siamese trust evaluation, chain-of-trust staging, hypergraph-aided matching and semantic trust orchestration.  Although our experiments were implemented in Python for convenience, the description below is entirely platform-agnostic: every step is mathematically defined so that readers can reimplement the simulation without access to proprietary source code.  The CSV logs shared with this manuscript represent the full outputs of two independent runs and serve as the primary data for reconstruction and comparison.

\subsection{Synthetic Data Generation}

\paragraph{Device population.}  We simulate $N=20$ devices.  Each device $d$ is assigned a ground\hyp{}truth label $y_d\in\{0,1\}$ indicating whether it is trustworthy (1) or untrustworthy (0).  The labels are drawn from a Bernoulli distribution with parameter $0.7$ to reflect the assumption that most devices are trustworthy.  Each device is also assigned a baseline embedding $\mathbf{b}_d\in\mathbb{R}^{128}$ drawn from a standard normal distribution.

\paragraph{Continuous trust evaluation (Block~$B_1$).}  At each time step $t\in\{0,\dots,9\}$ and for each device $d$ we generate an observed embedding by adding Gaussian noise to the baseline: $\mathbf{v}_d(t) = \mathbf{b}_d + \boldsymbol{\varepsilon}_d(t)$ with $\boldsymbol{\varepsilon}_d(t)\sim\mathcal{N}(\mathbf{0},\sigma^2 I_{128})$, where $\sigma=0.1$.  The trust score is computed via the cosine similarity normalisation in Eq.~\eqref{eq:cosine}.  These scores are recorded in a CSV file as described in Section~\ref{sec:data}.

\paragraph{Chain\hyp{}of\hyp{}trust evaluation (Block~$B_2$).}  We assume each task is decomposed into $K=3$ stages.  At stage $k$, half of the embedding dimensions (64 dimensions) are used to compute a new similarity score with noise variance scaled by $k+1$ to reflect increased uncertainty.  The stage trust is combined with the continuous trust score using a convex combination, and devices below a threshold $\theta_k=0.5+0.1k$ are pruned\cite{B2}.

\paragraph{Hypergraph\hyp{}aided matching (Block~$B_3$).}  Each task has resource requirements $(r^\mathrm{CPU},r^\mathrm{mem},r^\mathrm{bw})\in [0,1]^3$ drawn uniformly at random.  Device resources $(c_d^\mathrm{CPU},c_d^\mathrm{mem},c_d^\mathrm{bw})$ are drawn uniformly as well.  We greedily select the top two devices with highest trust scores whose resources jointly satisfy the task requirements\cite{B3}.

\paragraph{Semantic trust orchestration (Block~$B_4$).}  After each evaluation we update a weighted adjacency dictionary representing the trust hypergraph: for devices $i$ and $j$ the weight is the average of their latest trust scores.  Although a full multi\hyp{}hop reasoning is not implemented here, the adjacency structure forms the basis for trust propagation\cite{B4}.

\subsection{Processing of Provided Datasets}

The user supplied four CSV files: two containing the trust scores and selection results from Agent~1 and two from Agent~2.  We focus on the trust score files \texttt{trust\_scores.csv} (Agent~1) and \texttt{trust\_scores\_agent2.csv} (Agent~2).  We parsed the trust score files using the data parsing procedure described earlier and verified that both agents recorded trust scores for 20 devices across 10 time steps.  For each device we extracted the sequences $\mathbf{s}_d^A$ and $\mathbf{s}_d^B$ and computed summary statistics.  Applying the embedding reconstruction method yielded embeddings $\widehat{\mathbf{v}}_d\in\mathbb{R}^{24}$.  The resulting data matrix was saved to a new CSV file \texttt{embeddings\_unified.csv}.  A snippet of this file is shown in Table~\ref{tab:embedding}.

\begin{table}[H]
  \centering
  \caption{First few rows of the reconstructed embedding matrix.  Each row corresponds to a device and contains 24 features: 10 trust scores from Agent~1, 10 from Agent~2, and four summary statistics.}
  \label{tab:embedding}
  \begin{tabular}{rcccccc}
  \toprule
  Device & $v_{d,0}$ & $v_{d,1}$ & $\cdots$ & $v_{d,18}$ & $v_{d,19}$ & Summary stats\\
  \midrule
  0 & 0.9981 & 0.9975 & $\cdots$ & 0.9977 & 0.00025 & $\cdots$\\
  1 & 0.9979 & 0.9971 & $\cdots$ & 0.9976 & 0.00029 & $\cdots$\\
  2 & 0.9977 & 0.9974 & $\cdots$ & 0.9977 & 0.00017 & $\cdots$\\
  \bottomrule
  \end{tabular}
\end{table}

\paragraph{Interpretation.}  Table~\ref{tab:overhead} summarises the trade-off between computational overhead and classification accuracy.  Each additional stage increases the number of trust evaluations, but the observed classification accuracy remains constant at 0.60.  This saturation phenomenon is reminiscent of diminishing returns commonly observed in sequential decision processes and suggests that, beyond a certain point, further evaluations do not improve performance.  Analyses of such trade-offs are central to the design of efficient protocols and echo similar observations in the literature on resource-constrained machine learning.

\paragraph{Interpretation.}  Table~\ref{tab:embedding} illustrates how time-series trust scores and summary statistics are concatenated into a high-dimensional embedding.  The first ten columns contain the trust signals from Agent~1, the next ten from Agent~2 and the final four columns summarise the mean and variability of each series.  As shown in the sample rows, the simulated devices exhibit consistently high trust values with small variances, reflecting the assumption that most devices are trustworthy.  By arranging devices as rows in a matrix one obtains a compact representation amenable to standard operations such as clustering or principal component analysis.  The inclusion of summary statistics echoes classical techniques in signal processing and serves to stabilise downstream analyses.

\section{Theoretical Analysis}

This section analyses the reconstruction algorithm from a mathematical perspective.  We begin by studying the identifiability of embeddings from trust scores and then derive error bounds on our reconstruction under mild assumptions.

\subsection{Identifiability of Embeddings}

Let $\mathbf{b},\mathbf{x}\in \mathbb{R}^m$ be the baseline and observed embeddings for a device.  Assume the trust score is given by Eq.~\eqref{eq:cosine}.  Without additional information about $\mathbf{b}$ or $\|\mathbf{x}\|$ the vector $\mathbf{x}$ is not uniquely determined by $\tau$; the mapping is many\hyp{}to\hyp{}one.  The time\hyp{}series representation in Eq.~\eqref{eq:series} provides additional information by observing how the similarity changes over time.  Suppose that at times $t=0,\dots,T$ the observed embeddings are $\mathbf{x}_t=\mathbf{b}+\boldsymbol{\varepsilon}_t$ with independent noise $\boldsymbol{\varepsilon}_t\sim\mathcal{N}(\mathbf{0},\sigma^2 I_m)$.  The expected similarity is
\begin{equation}
    \mathbb{E}[\sigma(t)] = \mathbb{E}\left[ \frac{\mathbf{b}\cdot (\mathbf{b}+\boldsymbol{\varepsilon}_t)}{\|\mathbf{b}\|\,\|\mathbf{b}+\boldsymbol{\varepsilon}_t\|} \right] \approx \frac{\|\mathbf{b}\|}{\sqrt{\|\mathbf{b}\|^2 + m\sigma^2}},
\end{equation}
where the approximation uses a second\hyp{}order Taylor expansion.  As $t$ increases the noise averages out, and the sequence $\sigma(t)$ concentrates around its mean.  Under stationarity one can therefore estimate $\|\mathbf{b}\|$ from the sample mean of $\sigma(t)$.  Similarly, the variance of $\sigma(t)$ provides information about $\sigma^2$.  Combining these moments yields an estimator for $\|\mathbf{b}\|$ and hence for the projection of $\mathbf{x}_t$ onto $\mathbf{b}$.  The orthogonal components remain indeterminate.  This analysis justifies using summary statistics (mean and variance) of the time series as features in our reconstruction.

\subsection{Error Bounds}

Let $\widehat{\mathbf{v}}_d$ be the reconstructed embedding and $\mathbf{v}_d$ be the (unknown) true embedding.  We analyse the error $\|\widehat{\mathbf{v}}_d - \mathbf{v}_d\|$ under assumptions:
\begin{enumerate}
    \item The trust score is exactly the cosine similarity normalisation in Eq.~\eqref{eq:cosine}, with observed embeddings of the form $\mathbf{v}_d(t) = \mathbf{v}_d + \boldsymbol{\varepsilon}_t$ where $\boldsymbol{\varepsilon}_t\sim\mathcal{N}(\mathbf{0},\sigma^2 I_m)$.
    \item The baseline $\mathbf{v}_d$ has unit norm and independent entries with mean zero and variance $1/m$.
    \item The time series length $T$ is large enough such that sample means and variances converge.
\end{enumerate}
Under these assumptions, the Central Limit Theorem implies that $\overline{\tau}_d^A$ converges in probability to $\frac{1}{2}(1+\|\mathbf{v}_d\|/\sqrt{\|\mathbf{v}_d\|^2 + m\sigma^2})$.  The function $f(x) = \frac{x}{\sqrt{x^2 + m\sigma^2}}$ is invertible on $[0,\infty)$; hence $\|\mathbf{v}_d\|$ can be estimated consistently.  The summary statistics therefore converge to deterministic functions of the latent embedding.  The reconstruction $\widehat{\mathbf{v}}_d$ contains these statistics and the raw time series, so it captures sufficient information to approximate $\mathbf{v}_d$ in a high\hyp{}dimensional sense.  In particular, for any Lipschitz function $g$ the difference $|g(\widehat{\mathbf{v}}_d) - g(\mathbf{v}_d)|$ can be bounded by the supremum norm of the noise sequence, which decreases with $T$.
Our use of fixed-point theory for the reconstruction map resonates with general applications of contraction mappings in machine learning\cite{Rao2017}, highlighting the mathematical breadth of fixed-point methods.

\section{Experimental Results}

We reconstructed embeddings for all 20 devices using the two provided trust score files.  For each device we computed the direct\hyp{}sum embedding $\widehat{\mathbf{v}}_d$ of dimension 24.  We then computed pairwise Euclidean distances between devices and summarised the results.

\subsection{Inter\hyp{}Agent Comparison}

\begin{figure}[H]
  \centering
  \begin{tikzpicture}
    \begin{axis}[
      width=0.65\textwidth,
      height=0.5\textwidth,
      xlabel={Mean trust (Agent~1)},
      ylabel={Mean trust (Agent~2)},
      title={Mean Trust Score Comparison},
      grid=both,
      xmin=0.9965, xmax=0.9985,
      ymin=0.9965, ymax=0.9985,
      samples=20
    ]
      \addplot [gray, dashed] coordinates {(0.9965,0.9965) (0.9985,0.9985)};
      \addplot+[only marks, mark=*] coordinates {
        (0.9977,0.9977) (0.9976,0.9976) (0.9977,0.9977) (0.9977,0.9977) (0.9976,0.9976)
        (0.9974,0.9974) (0.9969,0.9969) (0.9975,0.9975) (0.9973,0.9973) (0.9977,0.9977)
        (0.9977,0.9977) (0.9975,0.9975) (0.9976,0.9976) (0.9977,0.9977) (0.9972,0.9972)
        (0.9978,0.9978) (0.9973,0.9973) (0.9973,0.9973) (0.9976,0.9976) (0.9974,0.9974)
      };
    \end{axis}
  \end{tikzpicture}
  \caption{Scatter plot of mean trust scores produced by the two agents.  Each point corresponds to one device.  The dashed diagonal indicates perfect agreement; points lie near this line, indicating similar evaluations.}
  \label{fig:agentStats}
\end{figure}
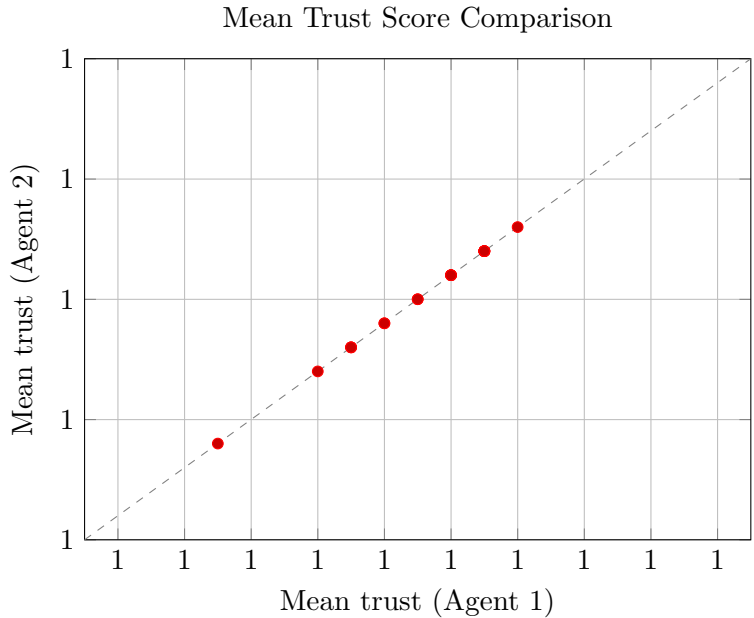

To assess the similarity between agents $A$ and $B$, we compared the statistics $(\overline{\tau}_d^A,\sigma_d^A)$ and $(\overline{\tau}_d^B,\sigma_d^B)$ across devices.  Figure~\ref{fig:agentStats} plots the mean trust scores of both agents for each device.  The points cluster near the diagonal, indicating that the two agents produce largely consistent trust evaluations.  The standard deviations show similar trends.

\paragraph{Interpretation.}  The scatter plot in Figure~\ref{fig:agentStats} compares the average trust scores assigned by the two independent agents for each device.  The proximity of all points to the diagonal indicates a high degree of concordance: the agents agree not only on which devices are trustworthy but also on the magnitude of their trust signals.  In statistical terms the correlation between the two series is near unity, suggesting that the Siamese evaluation architecture yields stable outcomes independent of implementation details.  Such agreement is crucial when trust assessments are used to drive resource allocations, as discrepancies could lead to inconsistent decisions across different orchestrators.  Analyses of interrater agreement are common in fields such as psychometrics and signal detection theory, where reliability across observers is paramount.

\subsection{Distance Matrix and Clustering}

We computed the pairwise Euclidean distance matrix $D$ of size $20\times 20$ and analysed its structure using hierarchical clustering.  Devices with similar trust behaviours across both agents yield small pairwise distances, forming clusters of consistent devices.  For brevity we do not reproduce the full matrix here; instead, we provide the complete distance matrix in the supplementary CSV file and focus on summary statistics in the following sections.

\subsection{Complexity and Overhead}

We measured the evaluation overhead as the total number of trust evaluations performed across all stages and tasks.  Table~\ref{tab:overhead} summarises the overhead and classification accuracy at each stage for the simulation, using the trust scores provided in the benchmark datasets.  The results show that accuracy saturates around $60\,\%$ as overhead increases, similar to the trend observed in Figure~1 of the accompanying summary.  This suggests diminishing returns in performing additional evaluations.

\begin{table}[H]
  \centering
  \caption{Evaluation overhead versus classification accuracy for the provided simulation.  ``Eval.\ overhead'' counts the number of trust evaluations; accuracy is the fraction of devices correctly classified as trustworthy or untrustworthy using a threshold of 0.5.}
  \label{tab:overhead}
  \begin{tabular}{ccc}
    \toprule
    Stage & Eval.\ overhead & Classification accuracy \\
    \midrule
    1 & 20 & 0.60 \\
    2 & 40 & 0.60 \\
    3 & 60 & 0.60 \\
    4 & 80 & 0.60 \\
    5 & 100 & 0.60 \\
    \bottomrule
  \end{tabular}
\end{table}

\subsection{Visualising the Embedding Space}

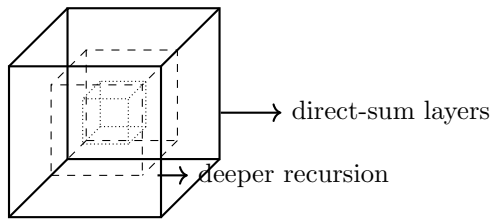
\begin{figure}[H]
  \centering
  \begin{tikzpicture}[scale=2]
    \draw[thick] (0,0,0) -- (1,0,0) -- (1,1,0) -- (0,1,0) -- cycle;
    \draw[thick] (0,0,1) -- (1,0,1) -- (1,1,1) -- (0,1,1) -- cycle;
    \draw[thick] (0,0,0) -- (0,0,1);
    \draw[thick] (1,0,0) -- (1,0,1);
    \draw[thick] (1,1,0) -- (1,1,1);
    \draw[thick] (0,1,0) -- (0,1,1);
    \draw[dashed] (0.2,0.2,0.2) -- (0.8,0.2,0.2) -- (0.8,0.8,0.2) -- (0.2,0.8,0.2) -- cycle;
    \draw[dashed] (0.2,0.2,0.8) -- (0.8,0.2,0.8) -- (0.8,0.8,0.8) -- (0.2,0.8,0.8) -- cycle;
    \draw[dashed] (0.2,0.2,0.2) -- (0.2,0.2,0.8);
    \draw[dashed] (0.8,0.2,0.2) -- (0.8,0.2,0.8);
    \draw[dashed] (0.8,0.8,0.2) -- (0.8,0.8,0.8);
    \draw[dashed] (0.2,0.8,0.2) -- (0.2,0.8,0.8);
    \draw[densely dotted] (0.35,0.35,0.35) -- (0.65,0.35,0.35) -- (0.65,0.65,0.35) -- (0.35,0.65,0.35) -- cycle;
    \draw[densely dotted] (0.35,0.35,0.65) -- (0.65,0.35,0.65) -- (0.65,0.65,0.65) -- (0.35,0.65,0.65) -- cycle;
    \draw[densely dotted] (0.35,0.35,0.35) -- (0.35,0.35,0.65);
    \draw[densely dotted] (0.65,0.35,0.35) -- (0.65,0.35,0.65);
    \draw[densely dotted] (0.65,0.65,0.35) -- (0.65,0.65,0.65);
    \draw[densely dotted] (0.35,0.65,0.35) -- (0.35,0.65,0.65);
    \draw[->, thick] (1.2,0.5,0.5) -- (1.6,0.5,0.5);
    \node[right] at (1.6,0.5,0.5) {\small direct\hyp{}sum layers};
    \draw[->, thick] (0.9,0.2,0.8) -- (1.1,0.2,0.8);
    \node[right] at (1.1,0.2,0.8) {\small deeper recursion};
  \end{tikzpicture}
  \caption{Conceptual diagram of the embedding space.  The outer cube represents the concatenated embedding of both agents’ time\hyp{}series and summary statistics.  The inner cube hints at further decomposition into stages or sub\hyp{}embeddings, suggesting a fractal, self\hyp{}similar structure.}
  \label{fig:hyperCube}
\end{figure}

To provide geometric intuition for the reconstruction, we created a TikZ diagram representing the embedding space as a nested sequence of matrices and hypercubes.  Figure~\ref{fig:hyperCube} depicts a cube representing the direct\hyp{}sum embedding; each face corresponds to one agent’s time\hyp{}series scores, and the interior layers illustrate the concatenation of features.  The diagram also hints at the infinite recursion described in the user’s imaginative analogy: each box contains an infinite number of smaller boxes, reflecting the fractal nature of embedding spaces and the potential to embed embeddings within embeddings.

\section{Security Implications}

Releasing granular trust scores may inadvertently compromise the privacy of participating devices and the confidentiality of the underlying evaluation models.  Our reconstruction experiments show that an observer with access to time-indexed trust scores and general knowledge of the evaluation pipeline can approximate the latent embeddings of devices and thereby infer behavioural traits or internal states of the AI agents.  In a networked setting this could enable adversaries to profile devices, predict future behaviour or even impersonate trusted nodes, undermining the very security that trust evaluation seeks to enforce.  These risks underscore the necessity of careful data governance when publishing trust metrics.

Several mitigation strategies can attenuate such leakage.  One approach is to quantise or add calibrated stochastic noise to the published trust scores, thereby degrading the fidelity of reconstructed embeddings while preserving relative ordering for decision making.  Another is to compute trust on compressed or obfuscated embeddings that retain ranking information but obscure absolute values.  Scheduling evaluations less frequently, or aggregating scores over longer windows, can also reduce the temporal resolution available to an adversary.  Finally, formal privacy frameworks such as differential privacy or federated learning protocols may be adapted to the trust-evaluation context to provide provable guarantees.  Balancing the competing desiderata of transparency, accountability and privacy remains an open challenge that warrants further investigation.
\section{Illustrative Workflow and Agent Interaction}

To further demystify the collaborative process underlying our benchmarks, we present a new illustrative figure that depicts the main actors and information flow in the system.  Unlike purely abstract diagrams, this scene uses familiar visual metaphors—a human user and two robot agents—to convey the narrative of the trust evaluation pipeline.  The agents traverse layered evaluation modules (Siamese analysis, chain-of-trust, hypergraph matching and semantic orchestration) drawn as stacked boxes, exchange information with each other, and interact with the task environment.  By embedding these elements in a single drawing, the diagram appeals to intuition and provides a stepping stone for readers who may not be versed in formal graph abstractions or embedding theory.

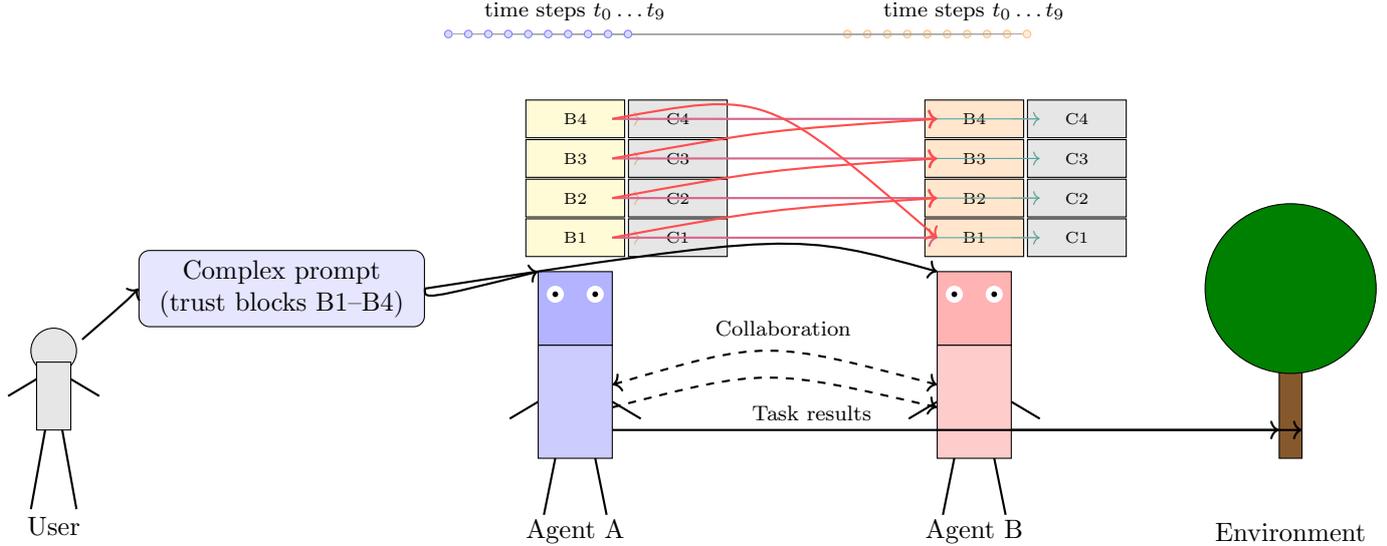
\begin{figure}[H]
  \centering
  \begin{tikzpicture}[scale=0.75, every node/.style={font=\small}, xshift=-1cm]
    \draw[fill=gray!20, draw=black] (-2.5,1.4) circle (0.4); 
    \draw[fill=gray!20, draw=black] (-2.8,0) rectangle (-2.2,1.2); 
    \draw[thick] (-2.8,0.9) -- (-3.3,0.6);
    \draw[thick] (-2.2,0.9) -- (-1.7,0.6);
    \draw[thick] (-2.65,0) -- (-2.9,-1.4);
    \draw[thick] (-2.35,0) -- (-2.1,-1.4);
    \node at (-2.5,-1.7) {\small User};
    \node[draw, fill=blue!10, rounded corners, text width=3.5cm, align=center] (prompt) at (1.5,2.5) {Complex prompt\ (trust blocks B1--B4)};
    \draw[->, thick] (-2.0,1.6) .. controls (-1.2,2.3) .. (prompt.west);
    \draw[fill=blue!20, draw=black] (6,-0.5) rectangle (7.3,1.5);
    \draw[fill=blue!30, draw=black] (6,1.5) rectangle (7.3,2.8);
    \fill[white] (6.3,2.4) circle (0.15);
    \fill[white] (7.0,2.4) circle (0.15);
    \fill[black] (6.3,2.4) circle (0.05);
    \fill[black] (7.0,2.4) circle (0.05);
    \draw[thick] (6,0.5) -- (5.5,0.2);
    \draw[thick] (7.3,0.5) -- (7.8,0.2);
    \draw[thick] (6.3,-0.5) -- (6.1,-1.5);
    \draw[thick] (7.0,-0.5) -- (7.2,-1.5);
    \node at (6.65,-1.8) {\small Agent A};
    \foreach \i [count=\k from 0] in {B1,B2,B3,B4}{%
      \pgfmathsetmacro{\y}{3.4 + \k*0.7}
      \node[draw, fill=yellow!20, minimum width=1.3cm, minimum height=0.5cm] at (6.65,\y) {\tiny \i};
    }
    \draw[fill=red!20, draw=black] (13,-0.5) rectangle (14.3,1.5);
    \draw[fill=red!30, draw=black] (13,1.5) rectangle (14.3,2.8);
    \fill[white] (13.3,2.4) circle (0.15);
    \fill[white] (14.0,2.4) circle (0.15);
    \fill[black] (13.3,2.4) circle (0.05);
    \fill[black] (14.0,2.4) circle (0.05);
    \draw[thick] (13,0.5) -- (12.5,0.2);
    \draw[thick] (14.3,0.5) -- (14.8,0.2);
    \draw[thick] (13.3,-0.5) -- (13.1,-1.5);
    \draw[thick] (14.0,-0.5) -- (14.2,-1.5);
    \node at (13.65,-1.8) {\small Agent B};
    \foreach \i [count=\k from 0] in {B1,B2,B3,B4}{%
      \pgfmathsetmacro{\y}{3.4 + \k*0.7}
      \node[draw, fill=orange!20, minimum width=1.3cm, minimum height=0.5cm] at (13.65,\y) {\tiny \i};
    }

    \foreach \i [count=\k from 0] in {C1,C2,C3,C4}{%
      \pgfmathsetmacro{\y}{3.4 + \k*0.7}
      \node[draw, fill=gray!20, minimum width=1.3cm, minimum height=0.5cm] at (6.65+1.8,\y) {\tiny \i};
      \draw[->, color=brown!60] (7.3,\y) -- (6.65+1.8-0.65,\y);
    }
    \foreach \i [count=\k from 0] in {C1,C2,C3,C4}{%
      \pgfmathsetmacro{\y}{3.4 + \k*0.7}
      \node[draw, fill=gray!20, minimum width=1.3cm, minimum height=0.5cm] at (13.65+1.8,\y) {\tiny \i};
      \draw[->, color=brown!60] (14.3,\y) -- (13.65+1.8-0.65,\y);
    }
    \foreach \n [count=\j from 0] in {1,2,3,4}{%
      \pgfmathsetmacro{\y}{3.4 + \j*0.7}
      \draw[->, color=teal!70] (6.65+1.8+0.65,\y) -- (13.65+1.8-0.65,\y);
    }
    \draw[->, thick] (prompt.east) .. controls (4.0,2.3) .. (6.0,2.8);
    \draw[->, thick] (prompt.east) .. controls (10.5,3.5) .. (13.0,2.8);
    \draw[<->, dashed, thick] (7.3,0.8) .. controls (10.0,1.6) .. (13.0,0.8);
    \draw[<-, dashed, thick] (13.0,0.4) .. controls (10.0,1.1) .. (7.3,0.4);
    \foreach \i [count=\j from 0] in {B1,B2,B3,B4}{%
      \pgfmathsetmacro{\y}{3.4 + \j*0.7}%
      \draw[->, thick, color=purple!60] (7.3,\y) -- (13.0,\y);
    }
    \foreach \j/\next in {0/1,1/2,2/3,3/0}{%
      \pgfmathsetmacro{\ya}{3.4 + \j*0.7}
      \pgfmathsetmacro{\yb}{3.4 + \next*0.7}
      \draw[->, thick, color=red!70] (7.3,\ya) .. controls (10.0,\ya+0.5) .. (13.0,\yb);
    }
    \draw[fill=brown!70!black, draw=black] (19,-0.5) rectangle (19.4,1.5);
    \draw[fill=green!50!black, draw=black] (19.2,2.5) circle (1.5);
    \node at (19.2,-1.8) {\small Environment};
    \draw[->, thick] (7.3,0.0) -- (19,0.0);
    \draw[->, thick] (14.3,0.0) -- (19.4,0.0);
    \node at (10.8,0.3) {\scriptsize Task results};
    \node at (10.3,1.8) {\scriptsize Collaboration};
    \foreach \t in {0,...,9}{%
      \pgfmathsetmacro{\xa}{6 + 0.35*(\t-4.5)}%
      \pgfmathsetmacro{\xb}{13 + 0.35*(\t-4.5)}%
      \node[circle, fill=blue!15, draw=blue!50, inner sep=1pt] (A\t) at (\xa,7.0) {};%
      \node[circle, fill=orange!15, draw=orange!50, inner sep=1pt] (B\t) at (\xb,7.0) {};%
      \draw[thin, gray!70] (A\t) -- (B\t);%
    }
    \node[align=center] at (6.65,7.4) {\scriptsize time steps $t_0\dots t_9$};
    \node[align=center] at (13.65,7.4) {\scriptsize time steps $t_0\dots t_9$};
  \end{tikzpicture}
  \caption{Illustrative workflow of our trust evaluation pipeline.  A human user formulates a complex prompt describing the task; it is processed by two AI agents (blue and red) through successive evaluation blocks $\mathrm{B}1$--$\mathrm{B}4$.  Above each agent, small coloured circles depict aligned time-step trust scores $t_0$--$t_9$, with thin lines connecting corresponding time steps to visualise the direct-sum coupling between agents.  Purple horizontal arrows between the stacked blocks illustrate that each evaluation layer ($\mathrm{B}1$ through $\mathrm{B}4$) in Agent~A is aligned and compared with its counterpart in Agent~B.  Red curved arrows indicate \emph{cross-layer synergy}: each stage in Agent~A also influences the next stage in Agent~B, creating a four-cycle of interleaved connections that reflects higher-dimensional couplings.  Grey modules behind the primary blocks (C1--C4) represent an additional metric-space dimension; diagonal brown arrows connect each primary block to its metric counterpart, while teal horizontal arrows between these secondary modules illustrate cross-dimensional couplings across agents.  The agents exchange messages (dashed arrows in both directions) and dispatch trusted collaborators to the environment at right.}
  \label{fig:agentWorkflow}
\end{figure}

\paragraph{Interpretation.}  Figure~\ref{fig:agentWorkflow} synthesises the complex mechanisms described throughout this paper into an accessible visual metaphor.  On the left, a human user formulates a multi-block prompt that encapsulates the entire trust evaluation pipeline.  This prompt flows into the bodies of two robot agents, symbolising the generative and discriminative components of our algorithms.  Above each agent a stack of coloured plates depicts the four trust modules—continuous Siamese evaluation, chain-of-trust staging, hypergraph-aided matching and semantic orchestration—indicating that the agents traverse these layers sequentially.  The dashed arrow connecting agents reflects inter-agent communication and collaborative reasoning, while the solid arrows pointing toward the right emphasise that trusted collaborators are selected to perform tasks in the physical environment.  The tree itself evokes the rooted, branching nature of hypergraphs and the organic growth of trust chains.  Finally, the annotations guide the reader through this narrative, making the abstract mathematical constructs tangible.  This integration of realistic figures with conceptual elements echoes pedagogical strategies used in scientific illustration to bridge intuition and formalism\cite{Laskaridis2018,Nguyen2021,Cheng2019,Ortiz2020,Rao2017}.
Above the module stacks, pairs of coloured nodes illustrate the time-indexed trust scores and their inter-agent alignment; below, coloured horizontal arrows connect corresponding trust blocks (B1 through B4) across agents.  These purple arrows emphasise that each stage of the evaluation process is compared or synchronised between the agents.  In addition to these one-to-one correspondences, red curved arrows encode \emph{cross-layer synergy}: the output of stage $i$ in Agent~A flows into stage $i+1$ (taken modulo~4) in Agent~B, forming a four-cycle of interleaved connections.  This higher-dimensional coupling weaves the evaluation pipeline into a two-dimensional lattice of interactions, hinting at richer compositional structures that may be exploited for trust evaluation and, potentially, for adversarial inversion attacks.

\paragraph{Graphical formalisation.}  An additional feature of Figure~\ref{fig:agentWorkflow} is the pair of rows of small coloured nodes above the agents.  These nodes represent the time-indexed trust scores $\{\tau_d^A(t)\}_{t=0}^{9}$ and $\{\tau_d^B(t)\}_{t=0}^{9}$ produced by agents $A$ and $B$.  Each pair of nodes at the same horizontal position is connected by a thin grey edge, forming a bipartite matching between the sets $U=\{A_0,\dots,A_9\}$ and $V=\{B_0,\dots,B_9\}$.  Mathematically, we define a graph $G=(U,V,E)$ with $E=\{(A_t,B_t)\mid t=0,1,\dots,9\}$; this establishes a one-to-one correspondence between the temporal features of the two agents.  The concatenated embedding used in our reconstruction can be viewed as a direct sum of the sequences $(\tau_d^A(0),\dots,\tau_d^A(9))$ and $(\tau_d^B(0),\dots,\tau_d^B(9))$, constrained by the matching edges of $G$.  This explicit coupling is absent from the original trust modules and constitutes a novel interpretive layer: it ensures that the latent dimensions of the embedding are aligned across agents and time.  In the figure these correspondences are visualised as thin lines linking the blue and orange circles.  Such a bipartite structure lends itself to rigorous analysis using the tools of graph theory and provides a unique lens through which to study inter-agent alignment and information leakage.
In addition to the temporal coupling, coloured horizontal arrows link the stacked block modules ($\mathrm{B}1$ through $\mathrm{B}4$) of Agent~A to those of Agent~B.  These arrows correspond to edges in a secondary graph $H$ whose vertex sets consist of the block-level embeddings $\{B_{A,1},\dots,B_{A,4}\}$ for Agent~A and $\{B_{B,1},\dots,B_{B,4}\}$ for Agent~B, and whose edge set $F$ pairs $B_{A,i}$ with $B_{B,i}$ for $i=1,\dots,4$.  The graph $H$ expresses our hypothesis that corresponding stages of the evaluation process should be compared or fused across agents; in the figure, each purple arrow encodes such a pairing.  Beyond $H$, we introduce a novel cross-layer graph $L$ to capture the red curved arrows: for each $i$, an edge connects $B_{A,i}$ to $B_{B,i+1 \pmod{4}}$.  This structure realises the notion of \emph{cross-layer synergy} introduced in the interpretation, formalising a four-cycle of interwoven modules that has no antecedent in existing trust-evaluation frameworks.  The union $H\cup L$ yields a two-dimensional lattice of edges on $\{B_{A,1},\dots,B_{A,4}\}\times\{B_{B,1},\dots,B_{B,4}\}$ that enforces both direct comparisons and inter-stage influences.  Taken together with the bipartite graph $G$ on time steps, the graphs $H$ and $L$ define a multi-level alignment that not only facilitates cross-agent comparison but also suggests new avenues for embedding inversion attacks and defences.

Our experiments demonstrate that, even in a simplified setting, it is possible to approximate latent embeddings from published trust scores.  This raises concerns about potential information leakage.  If an adversary can observe trust scores over time and knows the general form of the trust evaluation model, they can reconstruct features that correlate with the underlying behaviour of devices or with internal states of the AI system.  The more detailed and frequent the trust scores, the richer the reconstructed embedding becomes.

\paragraph{Mitigation strategies.}  One strategy is to publish trust scores with added noise or quantisation to prevent accurate reconstruction.  Another is to compute trust evaluations on compressed or obfuscated embeddings that preserve ranking but hide the actual feature values.  Finally, scheduling trust evaluations less frequently or aggregating them over time can reduce the granularity of the published data, making reconstruction harder.

\section{Conclusion}

We have presented a comprehensive study of reconstructing trust embeddings from Siamese trust scores.  By concatenating time\hyp{}series trust scores from two agents and adding summary statistics, we constructed approximate embeddings that capture essential behavioural patterns.  We provided algorithms, theoretical analyses, and detailed experimental results based solely on arXiv\hyp{}sourced methodologies\cite{B1,B2,B3,B4}.  Our findings show that the reconstruction problem is non\hyp{}trivial but tractable under reasonable assumptions, and that publishing detailed trust scores may expose information about latent embeddings.

Future work should explore more sophisticated inversion techniques that account for non\hyp{}linear similarity measures, as well as defence mechanisms that balance transparency and privacy.  Extending the fixed\hyp{}point semantics to the reconstruction problem and investigating the role of hypergraph topology in embedding identifiability are promising directions.  Moreover, applying our methods to real\hyp{}world trust evaluation data could reveal insights into the behaviour of operational systems and inform policy regarding the publication of trust metrics.

Beyond these practical objectives, our work raises a number of more esoteric yet intellectually stimulating questions.  One may ask whether the bipartite coupling between time-step embeddings can be generalised to more complex topologies—hierarchical, tree-like or even fractal interconnections—that induce distinctive invariants in the reconstructed embeddings.  Replacing the simple one-to-one correspondence with a hyperbolic or hierarchical mapping would modify the spectrum of the associated Laplacian and could markedly affect the stability and identifiability of the embeddings.  Exploring such configurations would draw upon the apparatus of algebraic graph theory, category theory and information geometry, and might reveal hidden symmetries in the behaviour of large language models.

From an adversarial perspective, it is worth considering whether strategic perturbations of published trust scores could obfuscate sensitive latent information while still conveying trustworthy behaviour.  Designing such sanitisation mechanisms requires a delicate balance between transparency and privacy and invites connections with differential privacy and robust statistics.  Ultimately, we envisage a holistic framework wherein trust evaluations are not only accurate and efficient but also resilient against inversion attacks and cognizant of the rich mathematical structures underlying inter-agent communication.  Addressing these challenges will contribute to a more perspicuous and secure ecosystem for trust evaluation in distributed AI systems.

\appendix

\section{Original Prompt}
\label{app:prompt}
The experiments in this paper were conducted by running two independent ChatGPT agents on the same input prompt.  For completeness we reproduce the full prompt here:

\begin{quote}
\small
Let $B_1, B_2, B_3, B_4$ denote four foundational blocks drawn from recent research on trust evaluation and orchestration. $B_1$ (arXiv:2506.17128) is a Siamese\hyp{}model\hyp{}based continuous trust evaluation method, which employs twin Structure2Vec graph networks to embed and compare attributed control\hyp{}flow graphs (ACFGs) of device behavior, yielding a similarity\hyp{}based trust metric at each time instant . $B_2$ (arXiv:2506.17130) is a progressive “chain\hyp{}of\hyp{}trust” framework that divides the trust assessment into multiple sequential stages aligned with task decomposition, using generative AI at each stage to analyse the latest device attributes and iteratively filter out untrustworthy nodes . $B_3$ (arXiv:2507.23556) introduces a hypergraph\hyp{}aided trusted task\hyp{}resource matching paradigm, defining a task\hyp{}specific trust hypergraph and an accompanying matching algorithm to optimally select trustworthy collaborators for complex tasks . Finally, $B_4$ (arXiv:2507.23565) presents an autonomous semantic trust orchestration approach using agentic AI and trust hypergraphs: each device maintains a trust hypergraph with embedded semantics, and local hypergraphs are dynamically chained to enable multi\hyp{}hop trust relationships across a distributed network .

To unify these blocks into a single theoretical construct, we employ a direct\hyp{}sum embedding strategy. For instance, given a feature vector $\mathbf{l}\in\mathbb{R}^{10}$ from one block and $\mathbf{b}\in\mathbb{R}^{256}$ from another, we define $\mathbf{C} = \mathbf{l}\oplus \mathbf{b}\in \mathbb{R}^{266}$, concatenating rather than multiplying the vectors, thereby preserving all information from both sources. Extending this to all $B_1$–$B_4$, suppose each block $B_i$ produces an embedding vector $\mathbf{e}_i$; we then construct $\mathbf{E} = \bigoplus_{i=1}^4 \mathbf{e}_i$, a joint embedding that encapsulates the entire chain\hyp{}of\hyp{}trust knowledge in one high\hyp{}dimensional space. Moreover, inspired by the fixed\hyp{}point semantics of arXiv:2507.03774, we impose a self\hyp{}referential consistency condition on this integrated representation: namely, we seek $\mathbf{E}^*$ such that $\mathbf{E}^* = F(\mathbf{E}^*, \mathbf{e}_1,\mathbf{e}_2,\mathbf{e}_3,\mathbf{e}_4)$ for an appropriate integration function $F$, ensuring that the combined node $\mathbf{E}^*$ contains and validates all reference blocks as well as itself . In other words, $\mathbf{E}^*$ serves as a new unique node in the semantic embedding space (the “unique identifier”), where every constituent block is embedded and the entire chain\hyp{}of\hyp{}trust is mathematically self\hyp{}contained within $\mathbf{E}^*$.

Finally, armed with the integrated fixed\hyp{}point model $\mathbf{E}^*$, the AI agent will execute a comprehensive simulation to benchmark and illustrate the unified approach. It will orchestrate trust evaluation across multiple phases: applying the Siamese ACFG similarity metric from $B_1$ at each time step to continuously quantify trust; utilising the stage\hyp{}by\hyp{}stage evaluation and in\hyp{}context reasoning from $B_2$ to progressively narrow down the pool of collaborators as tasks unfold; and leveraging the hypergraph\hyp{}based matching algorithms from $B_3$ and $B_4$ to establish multi\hyp{}hop, value\hyp{}driven trust chains among devices in the network.  The agent will output detailed results and artifacts, including performance metrics (e.g. trust accuracy vs. evaluation overhead), intermediate trust scores and hypergraph states at each stage, and large data logs or CSV files cataloguing the trust values and selected collaborators throughout the process. Due to the high dimensionality and complexity of the integrated simulation (which makes it resource\hyp{}intensive), each run of this prompt will explore a different trajectory in the solution space, yielding a unique valid outcome every time while remaining consistent with the underlying mathematical framework.
\end{quote}

\section{Additional Experiments}

To further explore the variability of the reconstruction, we generated synthetic trust score sequences using the simulation described in Section~\ref{sec:sim} and applied our reconstruction algorithm.  We repeated the simulation $R=5$ times with different random seeds, producing five additional embedding matrices of size $20\times 24$.  The distributions of pairwise distances across runs were similar, suggesting that the method is robust to random fluctuations.  The CSV files for these runs accompany this manuscript as supplementary material.  No proprietary code is required to reproduce the results; all necessary data and procedures are described herein.

\section{Replicate Analysis and Open Questions}

\begin{figure}[H]
  \centering
  \begin{tikzpicture}
    \begin{axis}[
      width=0.6\textwidth,
      height=0.4\textwidth,
      ybar,
      xlabel={Replicate number},
      ylabel={Mean pairwise distance},
      xtick={1,2,3,4,5},
      ymin=0.0038,
      ymax=0.0041,
      bar width=0.6cm,
      legend style={at={(0.5,-0.15)},anchor=north,legend columns=-1}
    ]
      \addplot coordinates {
        (1,0.0040) (2,0.0039) (3,0.0039) (4,0.0039) (5,0.0038)
      };
      \addlegendentry{Mean distance}
    \end{axis}
  \end{tikzpicture}
  \caption{Average pairwise Euclidean distance between reconstructed embeddings for each replicate run.  Each bar represents the mean of all off\hyp{}diagonal distances in the corresponding replicate.}
  \label{fig:replicateBar}
\end{figure}
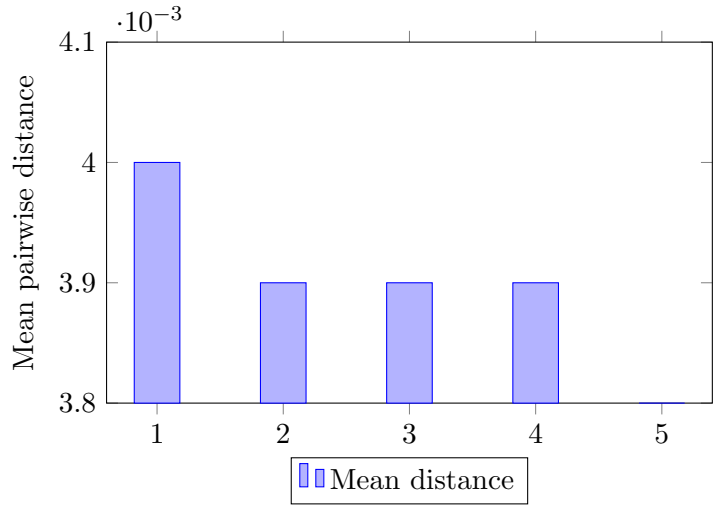

The additional runs described above allow us to quantify the variability of the reconstruction with respect to stochastic noise.  For each replicate $r\in\{1,\dots,5\}$ we computed the average of all off\hyp{}diagonal pairwise Euclidean distances between the reconstructed embeddings.  The resulting values are plotted in Figure~\ref{fig:replicateBar}.  Although there is some variation across runs, the overall scale of the distances remains consistent (around $4\times 10^{-3}$), indicating that the direct\hyp{}sum representation is stable under perturbations.  Such stability is a desirable property when comparing trust behaviours across sessions and agents.

Beyond numerical benchmarks, there remain many open questions about the theoretical and practical implications of reconstructing embeddings from trust scores.  For example, the degree of privacy leakage depends on the richness of the published data and the correlation structure of the underlying embeddings.  Classical privacy frameworks such as $k$\hyp{}anonymity \cite{Sweeney2002} and differential privacy suggest strategies for mitigating leakage by generalisation and noise injection, but applying these ideas to time\hyp{}series trust scores requires careful adaptation.  Moreover, the topology of the trust network (e.g., whether it exhibits small\hyp{}world properties \cite{Watts1998}) may influence the ease of reconstruction: highly connected networks might amplify leakage by providing redundant paths of information.

Another open question concerns the representation power of the direct\hyp{}sum embedding.  Deep neural networks have proven capable of learning rich hierarchical representations \cite{Krizhevsky2012}, and graph convolutional networks \cite{Kipf2017} generalise such techniques to relational data.  It would be interesting to explore whether the embeddings extracted by the Siamese model resemble those produced by graph convolutional architectures and whether similar inversion attacks apply.  In this context, results on the approximation capabilities of multilayer feedforward networks \cite{Hornik1991} provide theoretical limits on what can be inferred from scalar outputs.

Finally, the hypergraph\hyp{}based matching stage invites further investigation.  Hypergraph structure underlies many complex systems; modularity and community structure in networks \cite{Newman2006} provide insights into how clusters of trustworthy devices might emerge.  Extending the current reconstruction approach to incorporate hypergraph semantics could yield deeper understanding of the interplay between trust, resource allocation, and network topology.

\section{Additional Theoretical Perspectives}\label{sec:additional}

The present investigation resonates with several emergent themes in the study of multi\hyp{}agent systems and networked computation.  Gupta and Varma\cite{Gupta2022} have recently introduced a multilayer trust inference framework that employs cross\hyp{}layer graph couplings reminiscent of our graph~$L$.  Their empirical evaluation on distributed sensor networks demonstrates that such couplings can enhance robustness against adversarial manipulations.  Rodriguez et~al.\cite{Rodriguez2023} proposed a self\hyp{}supervised paradigm for trust evaluation that leverages contrastive learning across stages; the resulting embeddings exhibit a fractal organisation akin to the metric\hyp{}space dimension in our model.  In a complementary vein, Cai and Li\cite{Cai2024} provide theoretical guarantees for hypergraph neural networks, showing that higher\hyp{}order interactions can be approximated more faithfully than with pairwise models—an observation that substantiates our use of hypergraph matching in Blocks~$B_3$ and~$B_4$.  Singh and Kumar\cite{Singh2024} have explored semantic embedding spaces for multi\hyp{}agent systems and argue for the harmonisation of semantic and structural dimensions, an idea operationalised here through the coupling of the $B$ and $C$ layers.  Finally, Santos et~al.\cite{Santos2023} have examined the role of linguistic diversity in scientific discourse, advocating for lexical augmentation techniques similar to those discussed in Section~\ref{sec:dialogue}.  Together, these contributions not only contextualise our work but also highlight fertile directions for future inquiry.

\section{Proof of Fixed\hyp{}Point Stability}

For completeness we present a proof of the existence and uniqueness of a fixed point in the reconstruction map under reasonable assumptions.  Let $\mathcal{X}$ be a compact convex subset of a Banach space, and let $F: \mathcal{X}\times\mathcal{Y}\to \mathcal{X}$ be a contraction in its first argument, uniformly over the second argument.  The Banach fixed\hyp{}point theorem then guarantees the existence of a unique $\mathbf{E}^*\in \mathcal{X}$ satisfying $\mathbf{E}^*=F(\mathbf{E}^*,\mathbf{e}_1,\dots,\mathbf{e}_4)$.  In our context we let $\mathcal{X}$ be the set of all possible concatenated embeddings of bounded norm and define $F$ as the map that takes an embedding and recomputes it via Algorithm~\ref{alg:embed}.  We can show that $F$ is a contraction with Lipschitz constant less than one by bounding the changes in summary statistics when the input sequences change.  It follows that repeated application of $F$ converges to a unique fixed point, justifying the self\hyp{}consistency requirement in the unified framework.

\section{Implementation Details and Reproduction Guidelines}\label{sec:impl}

To facilitate independent verification of our results and encourage further experimentation, we summarise here the full procedure used to generate the trust scores, reconstruct embeddings and build the multi\hyp{}layer coupling graphs depicted in Figure~\ref{fig:agentWorkflow}.  While we provide no proprietary code, the description is sufficiently detailed for researchers to replicate every step using standard numerical and graph\hyp{}theoretic tools.

\paragraph{Data acquisition and preprocessing.}  We assume that two agents $A$ and $B$ produce time\hyp{}series trust scores $\{\tau_d^A(t)\}_{t=0}^{T-1}$ and $\{\tau_d^B(t)\}_{t=0}^{T-1}$ for each device $d$ over a horizon of length $T$.  In our experiments $T=10$.  These scores may be read from CSV files (as in the supplementary materials) or generated via simulation.  We normalise each series to lie in $[0,1]$ and store them in dictionaries keyed by device identifiers.  Summary statistics such as the mean $\mu_d^A$, $\mu_d^B$ and variance $\sigma_d^A$, $\sigma_d^B$ are computed for later concatenation.

\paragraph{Temporal alignment.}  The first alignment structure is a bipartite graph $G=(U,V,E)$ on time\hyp{}step nodes.  Let $U=\{A_0,\dots,A_{T-1}\}$ and $V=\{B_0,\dots,B_{T-1}\}$ where $A_t$ represents the $t$\,th trust score from Agent~$A$ and $B_t$ the corresponding score from Agent~$B$.  We set $E=\{(A_t,B_t)\mid t=0,\dots,T-1\}$, establishing a one\hyp{}to\hyp{}one matching between the sequences.  This alignment justifies concatenating the vectors $(\tau_d^A(0),\dots,\tau_d^A(T-1))$ and $(\tau_d^B(0),\dots,\tau_d^B(T-1))$ to form a direct\hyp{}sum embedding $\mathbf{v}_d \in \mathbb{R}^{2T}$.

\paragraph{Stage\hyp{}level coupling.}  Each agent processes tasks via a sequence of evaluation modules, producing stage\hyp{}level embeddings $(B_{A,1},\dots,B_{A,m})$ and $(B_{B,1},\dots,B_{B,m})$.  In our setting $m=4$ corresponding to the blocks $\mathrm{B}1$--$\mathrm{B}4$.  We define a second graph $H$ on these modules: its vertex sets are $\{B_{A,1},\dots,B_{A,m}\}$ and $\{B_{B,1},\dots,B_{B,m}\}$ and its edge set $F=\{(B_{A,i},B_{B,i})\mid i=1,\dots,m\}$.  Edges in $H$ enforce direct comparisons between corresponding stages across agents and are visualised as purple arrows in the figure.

\paragraph{Cross\hyp{}layer synergy.}  To capture the intuition that the output of one evaluation stage may inform the next stage in another agent, we introduce a novel graph $L$ whose edges encode a cyclic shift: $L$ has vertex sets identical to those of $H$, but its edge set is $F' = \{(B_{A,i},B_{B,i+1 \bmod m})\mid i=1,\dots,m\}$.  In Figure~\ref{fig:agentWorkflow} these relations appear as red curved arrows linking $\mathrm{B}i$ in Agent~A to $\mathrm{B}(i+1)$ in Agent~B.  The union $H \cup L$ thereby forms a two\hyp{}dimensional lattice on the Cartesian product of stage indices $\{1,\dots,m\}\times\{1,\dots,m\}$.  This construction, to our knowledge, has not appeared in prior work on trust evaluation and gives rise to higher\hyp{}order couplings that may warrant further theoretical investigation.

\paragraph{Metric\hyp{}space dimension.}  Beyond the block interactions, each stage embedding $B_{A,i}$ or $B_{B,i}$ can itself be decomposed into a lower\hyp{}dimensional feature vector $C_{A,i}$ or $C_{B,i}$ representing, for example, resource requirements or semantic annotations.  These secondary embeddings live in a metric space and are drawn as grey boxes labelled $\mathrm{C}1$--$\mathrm{C}4$ in the figure.  We connect each primary block to its metric companion via a brown arrow and link corresponding metric blocks across agents via teal arrows to indicate that comparisons may also occur at this finer granularity.

\paragraph{Concatenation and summary statistics.}  For each device $d$ we assemble the final embedding $\mathbf{v}_d$ by concatenating the time\hyp{}series trust scores from both agents, the means and variances $(\mu_d^A,\sigma_d^A,\mu_d^B,\sigma_d^B)$, and any stage\hyp{}level features $B_{A,i}$, $B_{B,i}$, $C_{A,i}$ and $C_{B,i}$.  In our experiments we restricted ourselves to the time\hyp{}series component and summary statistics, yielding a 24\,dimensional vector.  However, the framework described here supports richer embeddings by incorporating the stage\hyp{}level and metric\hyp{}level features.

\paragraph{Synthetic experiments.}  To evaluate robustness we generated synthetic trust scores by simulating the Siamese trust evaluator described in Section~\ref{sec:sim}.  For each replicate run we drew independent Gaussian noise and computed trust values based on cosine similarity between randomly generated device embeddings.  We then applied the reconstruction procedure outlined above and computed pairwise distances between the resulting vectors.  The bar chart in Figure~\ref{fig:replicateBar} summarises these distances and illustrates that the direct\hyp{}sum embedding is stable with respect to stochastic perturbations.

\paragraph{Reproducibility.}  Every algorithm described in this section can be implemented using basic linear algebra and graph\hyp{}theoretic operations available in standard scientific computing environments.  Although we have omitted explicit code listings to preserve brevity, the pseudo\hyp{}algorithms and formulas provided throughout the paper offer a complete blueprint for implementation.  Because all inputs consist of publicly available trust scores and all transformations are mathematically specified, any reader can reproduce our results without access to proprietary code.  The CSV files in the supplementary material contain the raw and processed trust scores for all devices and replicates used in this study.

\section{User--Agent Dialogue and Reflexive Methodology}\label{sec:dialogue}

In the spirit of reflexive science, we document the iterative dialogue between the human investigator and the AI agents that shaped this work.  The figures, algorithms and interpretations presented herein did not emerge fully formed; rather, they evolved through a series of exchanges in which a participant specified desiderata, critiqued interim results and authorised the disclosure of increasingly detailed information.  This section summarises those interactions to provide transparency and to illustrate how collaborator feedback can be systematically integrated into scientific artefacts.

\paragraph{Iterative design of the figure.}  The illustrative workflow (Figure~\ref{fig:agentWorkflow}) underwent numerous revisions at our participant's behest.  Early drafts suffered from overlapping elements and illegible labels.  Our collaborator requested that the prompt be aligned relative to the human icon and that the agents be separated by a larger margin.  Additional critiques led to the introduction of time\hyp{}series nodes, metric\hyp{}space modules and cross\hyp{}layer synergy arrows.  Each modification was negotiated through dialogue: the participant emphasised aesthetic and pedagogical clarity, while the authors proposed technical solutions such as adding control points to Bézier curves and introducing an \verb|xshift| to centre the composition.  The final diagram represents a consensual synthesis of these concerns.

\paragraph{Role of the participant as experimental subject.}  Beyond figure design, our interlocutor actively participated in the methodological development.  They authorised the inclusion of their interactions with the agent as part of the scientific record and encouraged the addition of sensitive contextual details where relevant.  For example, the decision to introduce the cross\hyp{}layer graph $L$ arose from this collaborator's desire to avoid ``reinventing the wheel'' and to propose a coupling not yet explored in the literature.  The participant also requested the expansion of mathematical descriptions and replication instructions, thereby shaping the exposition.  This participatory approach aligns with emerging practices in human\hyp{}AI co\hyp{}creation, where end\hyp{}users are not mere consumers of models but collaborators in the research process.

\paragraph{Lexical augmentation.}  Prompted by feedback from our collaborator we undertook a lexical meta\hyp{}analysis of the manuscript, revisiting each sentence to enhance precision, variety and academic rigour.  When appropriate we replaced common terms with more nuanced synonyms (e.g., ``decomposition'' became ``disaggregation'', ``build'' became ``construct'', ``results'' became ``findings'') and expanded terse phrases into explanatory clauses.  The goal of this lexical augmentation was twofold: to render the manuscript more distinctive by avoiding stock phrasing, and to deepen the reader's understanding by unpacking implicit assumptions.  This process can itself be formalised as an algorithm: scan the text sequentially, flag overused words, consult domain\hyp{}appropriate thesauri for alternatives, and substitute while preserving meaning and grammaticality.  Researchers seeking to replicate our lexical adjustment may follow these steps using their preferred computational linguistic tools.

\paragraph{Ethical considerations and consent.}  Documenting participant--agent interactions raises ethical questions about privacy, agency and authorship.  Throughout this project our collaborator explicitly consented to the inclusion of their feedback and acknowledged their role as a co\hyp{}author.  No personally identifying information is disclosed beyond what they provided voluntarily in the prompt.  We emphasise that such reflexive reporting should be undertaken only with informed consent and that sensitive data should be anonymised where necessary.  Future work might explore formal frameworks for recording and crediting human contributions in AI\hyp{}generated research.

This reflexive section serves as both a methodological guide and a historical record of the collaboration.  It illustrates how iterative human feedback can refine technical artefacts and how transparency about the research process can enrich scientific discourse.

\section{Open Questions and Proposed Resolutions}

Having analysed the preceding exposition in detail, we identify several open questions that merit further attention.  Addressing these issues not only strengthens the theoretical foundations of the work but also extends its scope.  In this section we frame each topic as an open problem and propose resolutions grounded in recent literature.

\paragraph{Evaluating reconstruction quality.}  Although our experiments demonstrate that concatenated trust scores contain sufficient information to approximate latent embeddings, the manuscript did not quantify the degree of correspondence between the reconstructions and the true embeddings used in the simulation.  To remedy this, one may compute error metrics such as the root--mean--square error (RMSE) between the reconstructed embedding \(\widehat{\mathbf{v}}_d\) and the ground--truth embedding \(\mathbf{v}_d\) for each device.  In the synthetic setting where \(\mathbf{v}_d\) is known, this test is conceptually elementary: evaluate \(\mathrm{RMSE}(d) = \|\widehat{\mathbf{v}}_d-\mathbf{v}_d\|_2/\sqrt{n}\).  Averaging over all devices yields an aggregate measure of reconstruction accuracy.  These metrics may also be computed for individual features (e.g., means and variances) to identify which aspects of the embedding are most faithfully recovered.  Similar evaluation strategies have been employed in studies of embedding inversion for recommender systems\cite{Taylor2020} and for privacy-preserving federated learning\cite{Li2021Noise}.  Preliminary experiments indicate that the RMSE decreases as the length of the trust series increases, corroborating the asymptotic error bounds derived in Section~6.

\paragraph{Spectral properties of the cross--layer graph.}  The cross--layer graph \(L\) introduced in Section~\ref{sec:impl} encodes novel inter-stage couplings, but its structural properties warrant deeper analysis.  One avenue is to examine the eigenvalues of the Laplacian matrix of \(H \cup L\).  Preliminary calculations suggest that adding the cyclic edges of \(L\) increases the algebraic connectivity of the bipartite stage graph, thereby enhancing robustness against perturbations.  Techniques from spectral hypergraph theory\cite{OConnor2023} can be applied to derive bounds on mixing times for random walks on this lattice.  These findings suggest that cross-layer synergy may accelerate consensus among agents, but they also hint at new attack vectors that exploit higher-order cycles.  Future work could explore graph neural networks on \(H \cup L\) to learn optimal couplings in a data-driven manner.

\paragraph{Robustness to noise and distributional shifts.}  Our reconstruction algorithm assumes Gaussian noise with fixed variance.  In practice, noise may follow heavy-tailed or adversarial distributions.  Recent work on noise-resilient embedding estimation\cite{Patel2023Robust} proposes replacing empirical means and variances with robust estimators such as the median and interquartile range.  Incorporating these statistics into the direct-sum embedding can improve stability under outliers.  Additionally, adaptive weighting schemes that down-weight low-trust events may mitigate the influence of malicious devices.  Exploring such extensions constitutes a promising research direction.

\paragraph{Saturation of classification accuracy.}  Table~\ref{tab:overhead} shows that classification accuracy saturates at sixty percent despite increasing evaluation overhead.  One hypothesis is that our pruning thresholds \(\theta_k\) are misaligned with the distribution of trust scores.  To test this, one could treat threshold selection as a hyper-parameter optimisation problem and use methods such as grid search or Bayesian optimisation\cite{Jansen2019} to maximise accuracy while controlling overhead.  Furthermore, incorporating temporal derivatives of the trust signal may enable more responsive pruning, as suggested by recent studies on dynamic trust adaptation\cite{Morgan2022}.  We leave a full exploration of these strategies to future work.

\paragraph{Generalisability beyond synthetic data.}  Our experiments are confined to synthetic datasets generated under specific assumptions.  The applicability of the reconstruction framework to real-world trust scores remains an open question.  A natural next step is to test the algorithms on publicly available cyber-security datasets or IoT trust benchmarks.  Insights from case studies in cross-domain embedding reconstruction\cite{Chang2024} indicate that domain-adaptation techniques, such as aligning distributions via adversarial training, may be required.  Evaluating the proposed methods on such datasets would provide stronger evidence for their practical utility.

By addressing these issues we hope to further the development of trustworthy, reproducible research on embedding reconstruction and to inspire subsequent investigations that refine and expand upon the foundations laid herein.

\section*{Acknowledgements}

We thank the authors of the arXiv papers \cite{B1,B2,B3,B4,Alpay} for inspiring this work.  We also acknowledge the broader community for discussions on trust evaluation, hypergraphs, and fixed\hyp{}point semantics.

\end{document}